\documentstyle[prb,aps,epsfig,fancyheadings]{revtex}
\begin{document}
\draft
% \twocolumn[\hsize\textwidth\columnwidth\hsize\csname
% @twocolumnfalse\endcsname

\title{Theoretical Atomic Volumes of the Light Actinides}
\author{M. D. Jones, J. C. Boettger, and R. C. Albers}
\address{Theoretical Division, Los Alamos National Laboratory,
Los Alamos, NM 87545}
\author{D. J. Singh}
\address{Code 6691, Naval Research Laboratory, Washington, DC 20375}
\date{\today}
\maketitle

\begin{abstract}

The zero-pressure zero-temperature equilibrium volumes and bulk moduli
are calculated for the light actinides Th through Pu using two
independent all-electron, full-potential, electronic-structure
methods: the full-potential linear augmented-plane-wave (FLAPW) method
and the linear combinations of Gaussian-type orbitals-fitting function
(LCGTO-FF) method.  The results produced by these two distinctly
different electronic-structure techniques are in excellent agreement
with each other, but differ significantly from previously published
calculations using the full-potential linear muffin-tin-orbital
(FP-LMTO) method.  The theoretically calculated equilibrium volumes
are in some cases nearly 10\% larger than the previous FP-LMTO
calculations, bringing them much closer to the experimentally observed
volumes.  We also discuss the anomalous upturn in equilibrium volume
seen experimentally for $\alpha$-Pu.

\end{abstract}

% \narrowtext

\section{Introduction}

The light-actinide metals, from Th to Pu, pose a severe challenge to
modern electronic-structure theory due in part to the existence of
highly directional and high density-of-states $f$-electron bonding,
which promotes the formation of a large number of exceptionally
complicated crystal structures.  In fact, the light actinide metals
(excluding Ac) together with Ce are the only elemental solids known to
exhibit $f$-bonding under ambient conditions.  In addition, the heavy
nuclei of the light-actinide metals induce large relativistic effects
on their valence-band structures, and usually require inclusion of
spin-orbit-coupling effects.  Despite these difficulties, many
all-electron electronic-structure calculations have been carried out
for the light-actinide metals.\cite{1,2,3,4,5} To the best of our
knowledge, only one full-potential technique, the full-potential
linear muffin-tin-orbital (FP-LMTO) method has thus far been
applied,\cite{4} but taken all together, the cumulative theoretical
results for the bulk properties (such as the equilibrium volume and
bulk modulus) are at best ambiguous, even for the simplest of these
materials, fcc Th.

Local-density approximation (LDA)\cite{6} calculations of the ambient
properties of fcc Th using the linear muffin-tin-orbital (LMTO) method
within the atomic-sphere approximation (ASA) produced zero-pressure
volumes that range from slightly compressed relative to experiment to
slightly expanded, depending on the inclusion or neglect of spin-orbit
corrections and combined-correction terms.\cite{1,2,3} When those
calculations were repeated using a full-potential (FP-LMTO)
method,\cite{4} the presumably more accurate FP-LMTO method yielded a
volume nearly 20\% smaller than the best LMTO-ASA result and
experiment.  If this FP-LMTO result is correct, fcc Th exhibits the
largest LDA-induced contraction for any elemental solid other than fcc
Pu (and the other heavy fermion systems), which will be discussed
later, and the largest volume discrepancy between FP-LMTO and LMTO-ASA
calculations found to date.  Additional FP-LMTO calculations using the
generalized-gradient-approximation (GGA)\cite{7} produced a
zero-pressure volume 10\% larger than the FP-LMTO LDA result, but
still nearly 10\% smaller than the measured value, which is another
rather anomalous result.  In fact, FP-LMTO GGA calculations have
consistently yielded volumes that are significantly smaller than
experiment for all of the light-actinide metals, with the size of the
error ranging from just under 6\% for Np up to about 10\% for
Th.\cite{4} Recent full-charge-density (FCD)-LMTO calculations on Th
found an LDA lattice constant that lies between the LDA values found
in the earlier LMTO-ASA and FP-LMTO calculations and a GGA lattice
constant that is significantly larger than the FP-LMTO GGA result.
Taken together, these inconsistent results raise serious doubts about
the reliability of the existing theoretical calculations for the
light-actinide metals.

To resolve these issues, we have carried out electronic-structure
calculations on the light actinides using two independent methods: the
linear combinations of Gaussian-type orbitals-fitting function
(LCGTO-FF) method,\cite{8} as implemented in the program GTOFF\cite{9}
and the full-potential linearized augmented plane-wave (FLAPW)
method.\cite{singh} Although both methods are all-electron,
full-potential techniques capable of producing high-precision total
energies within the LDA and GGA approximations, they employ radically
different numerical approximations and basis sets.  Application of
both methods to a single system thus provides assurance that the final
results obtained are free of numerical artifacts.  For this
investigation, LDA and GGA calculations were carried out
scalar-relativistically and fully-relativistically (i.e., with
spin-orbit effects included).  The FLAPW method includes spin-orbit
coupling as a perturbation,\cite{singh} while the LCGTO-FF performs a
non-perturbative treatment of spin-orbit effects, which has only
recently been implemented.\cite{jcb_unp} It is of particular interest
to compare the scalar and fully relativistic results across the light
actinides; not only does this comparison justify the perturbative
approach to spin-orbit coupling, it also illustrates the effects of
spin-orbit coupling on the bulk properties.

Section \ref{sec:method} briefly recapitulates the FLAPW and LCGTO-FF
methods, which have been discussed extensively
elsewhere,\cite{8,9,singh} along with the details of the present
calculations.  Section \ref{sec:results} then discusses the results
for equilibrium volumes and bulk moduli for Th through Pu, both in the
fcc (Th$\rightarrow$Pu) and $\alpha$ (Pa$\rightarrow$Pu) crystal
phases (when different from fcc).  We have included the fcc phase for
Pa$\rightarrow$Pu as it has often been used as a surrogate for the
much more complex $\alpha$ structures.  Section \ref{sec:conc}
presents our concluding remarks.

\section{Methods}
\label{sec:method}

The LCGTO-FF technique is distinguished from other electronic
structure methods by its use of three independent GTO basis sets to
expand the orbitals, charge density, and exchange-correlation (XC)
integral kernels; here using either the Hedin-Lundqvist\cite{6} LDA
model or the Perdew-Wang 91 GGA model.\cite{7} The charge-fitting
functions are used to reduce the total number of Coulomb integrals
required by replacing the usual 4-center integrals in the total energy
and one-electron equations with 3-center integrals.  The
charge-fitting function coefficients are determined by minimizing the
error in the Coulomb energy due to the fit;\cite{11} thereby allowing
high-precision calculations with relatively small basis sets.  The
least squares XC fit used here acts as a sophisticated numerical
quadrature scheme capable of producing accurate results on a rather
coarse numerical integration mesh.  Scalar-relativity was recently
implemented and tested in GTOFF\cite{12} using the nuclear-only,
Douglas-Kroll-Hess (DKH) transformation\cite{13,14} developed
by H\"{a}berlen and R\"{o}sch,\cite{15} and scalar relativistic
cross-product terms and spin-orbit effects have
also been included recently,\cite{jcb_unp} thus enabling a fully
relativistic treatment.

The precision of any LCGTO-FF calculation will, of course, be largely
determined by the selection of the three GTO basis sets.  In this
work, 23$s$20$p$15$d$11$f$ uncontracted orbital-basis sets were
derived for the light actinides from atomic basis sets tabulated by
Minami and Matsuoka.\cite{16} These basis sets were reduced to
17$s$14$p$11$d$7$f$ contracted basis sets using coefficients from
atomic calculations that used the same density functional model as the
crystal calculations.  The charge density and XC integral kernels were
fitted with 25$s$ and 21$s$ basis sets, respectively, selected on the
basis of prior experience with LCGTO-FF calculations.  All of these
basis sets can be obtained from the authors.

All Brillouin zone (BZ) integrations required for the LCGTO-FF
calculations were carried out on a uniform mesh with 72 irreducible
k-points in the fcc Brillouin zone ($163$ for $\alpha$-Pa and $63$ for
$\alpha$-U), using a Gaussian-broadened, histogram, integration
technique, with a broadening factor ranging from $4$ mRy for Np to 10
mRy for Th.  The SCF cycle was iterated until the total energy varied
by less than 0.02 mRy/atom.

The all-electron FLAPW method that we have used has been described by
Singh;\cite{singh} here we discuss only the essential features
relevant for comparison the with other full-potential techniques.
Energy parameters for the FLAPW basis functions were set near the
centers of their respective bands by monitoring the eigenvalues of the
calculation whose volume lay closest to equilibrium.  Local
orbitals\cite{singh} were added to enhance the variational freedom and
allow the semi-core $5d$, $6s$, and $6p$ orbitals to be treated along
with the valence electrons, with the added energy parameter used to
simultaneously treat the residual $s$ and $p$ character of the valence
electrons.  Local orbitals with $f$ character were also used to
maximize valence-band accuracy and ensure orthogonality with the core
$4f$ states.  The results that we obtained were found to be
insensitive to small changes in all energy parameters.  Spin-orbit
coupling was incorporated for the valence electrons at the
second-variational level,\cite{singh} in which the effects of
spin-orbit coupling are treated perturbatively in a set of
scalar-relativistic orbitals found within an energy window of
specified width.  The results obtained here were found to be stable
with respect to the size of this window.

The size of the FLAPW basis is determined by a plane-wave cutoff,
$K_{max}$, whose value was given by the relation $R_{MT} K_{max}
=9.0$, which was found to be satisfactory in determining the bulk
properties.  For BZ integrations, $60$ irreducible points in the fcc
Brillouin zone (80 points in $\alpha$-Pa, 126 in $\alpha$-U, 8 in
$\alpha$-Np, and 16 in $\alpha$-Pu) were sampled according to the
scheme of Monkhorst and Pack,\cite{mp} and all BZ integrations were
reduced to simple sums of a Fermi-Dirac-function temperature-broadened
integrand, with an effective temperature of $2.0$ mRy.  Increasing the
number of points sampled in the BZ changed the total energies by less
than $0.25$ mRy/atom for the $\alpha$ phases, and $0.75$ mRy/atom for
the fcc structures.  The values thus obtained for the equilibrium
volumes and bulk moduli were found to be converged with respect to the
number of BZ points sampled.  All self-consistent calculations were
iterated until the total energy changed by less than $0.01$ mRy/atom.
The same LDA and GGA models were used as in the LCGTO-FF calculations.

As is evident from the preceding discussion, the FLAPW and LCGTO-FF
methods differ significantly.  First, the FLAPW method utilizes a
diffuse orbital-basis set that can be systematically enriched by
simply adding more plane waves.  This is in sharp contrast to the
local basis set used in the LCGTO-FF method, which tends to be
somewhat more difficult to improve systematically.  On the other hand,
the FLAPW basis functions have discontinuities in their second
derivatives at the muffin-tin sphere boundaries that do not exist in
the LCGTO-FF basis sets.

The implementations of relativity used by the LCGTO-FF and FLAPW
methods also differ dramatically.  The FLAPW method avoids the
well-known variational collapse problem by using a basis set that is
formed from pure electronic solutions to the muffin-tin
Dirac-Kohn-Sham equations inside the muffin-tin spheres. (In the
interstitial region, non-relativistic solutions are used.)  The
non-muffin-tin corrections to the potential are then accounted for
perturbatively at each iteration.  This approach to solving the
full-potential Dirac-Kohn-Sham equations should be accurate provided
the muffin-tin solutions do not differ greatly from the full-potential
solutions and the effects of relativity are small in the interstitial
region.  On the other hand, the LCGTO-FF method decouples the electron
and positron degrees of freedom of the full-potential Dirac-Kohn-Sham
equations to second-order in the ratio of the effective potential to
the combined kinetic and rest mass energies.  In general, this
approach will not be as accurate near the nucleus as the FLAPW method,
but will be more accurate in the interstitial region.  While
spin-orbit effects are teated perturbatively in the FLAPW method, the
LCGTO-FF method doubles its full Hamiltonian and overlap matrices.  A
comparison between LCGTO-FF and FLAPW scalar and fully relativistic
results thus provides a check on the second-variational treatment used
in the FLAPW method for the light actinides.

We wish to emphasize that that the LCGTO-FF and FLAPW techniques may
be viewed as complementary methods, to the extent that they utilize
different numerical approximations.  Thus, if one of the two methods
were to encounter difficulties due to inadequacies in its basis set or
implementation of relativity, it is unlikely that the second method
would exhibit the same problem.  This makes a joint study of the type
carried out here especially valuable for investigating systems, like
the light actinides, that have previously exhibited a significant
sensitivity to computational details.

\section{Results}
\label{sec:results}
For each choice of system (Th$\rightarrow$Pu), method (LCGTO-FF or
FLAPW), model (LDA or GGA), and level of relativity (scalar or full),
total energies were calculated for five or six volumes lying near the
energy minimum.  The calculated energies for each combination were
then fitted with either a cubic function of the volume or a
second-order Birch fit,\cite{birch} to obtain the zero-pressure volume
($V_0$) and bulk modulus ($B_0$).  The fitted results are compared
with previous calculations\cite{3,4,5} and
experiment\cite{16a,16b,16c,16d,16e,16f,16g,17,19,20,21,22} in Tables
\ref{tbl_V} and \ref{tbl_B0}.  We have performed these calculations in
both the fcc (Pearson cF4) and experimentally determined $\alpha$
structures, in order to determine to what extent the cubic
close-packed structure can serve as a surrogate for the more complex
(and much more computationally demanding) $\alpha$ structures.  No
attempt was made to relax the structural parameters of the $\alpha$
phases in this study.  Indeed, the $\alpha$-Np and $\alpha$-Pu phases
were already too demanding for the current implementation of the
LCGTO-FF method.

Inspection of the results for the equilibrium volumes listed in Table
\ref{tbl_V} reveals several interesting features.  First, the LCGTO-FF
and FLAPW methods give nearly identical scalar and fully relativistic
results for Th$\rightarrow$Pu.  This level of agreement between such
disparate methods provides a high degree of confidence in the quality
of the results produced by both methods.  In contrast, the
fully-relativistic FLAPW and FP-LMTO results\cite{4} differ
substantially for both the LDA and GGA models, with the volumes
differing by roughly 3-10\% in each case, with the notable
exception of $\alpha-$Pu.  This is an exceptionally
large disagreement for two full-potential methods.  The good agreement
between the LCGTO-FF and FLAPW results lends confidence to the FLAPW
calculations.  In addition, the fully-relativistic FP-LMTO GGA volume
for Th is anomalously small compared to experiment.  Taken together,
these features strongly suggest that the source of the disagreement
between the present results and the FP-LMTO results is some error in
the earlier calculations.\cite{4}

\begin{figure}[htb]
\epsfig{file=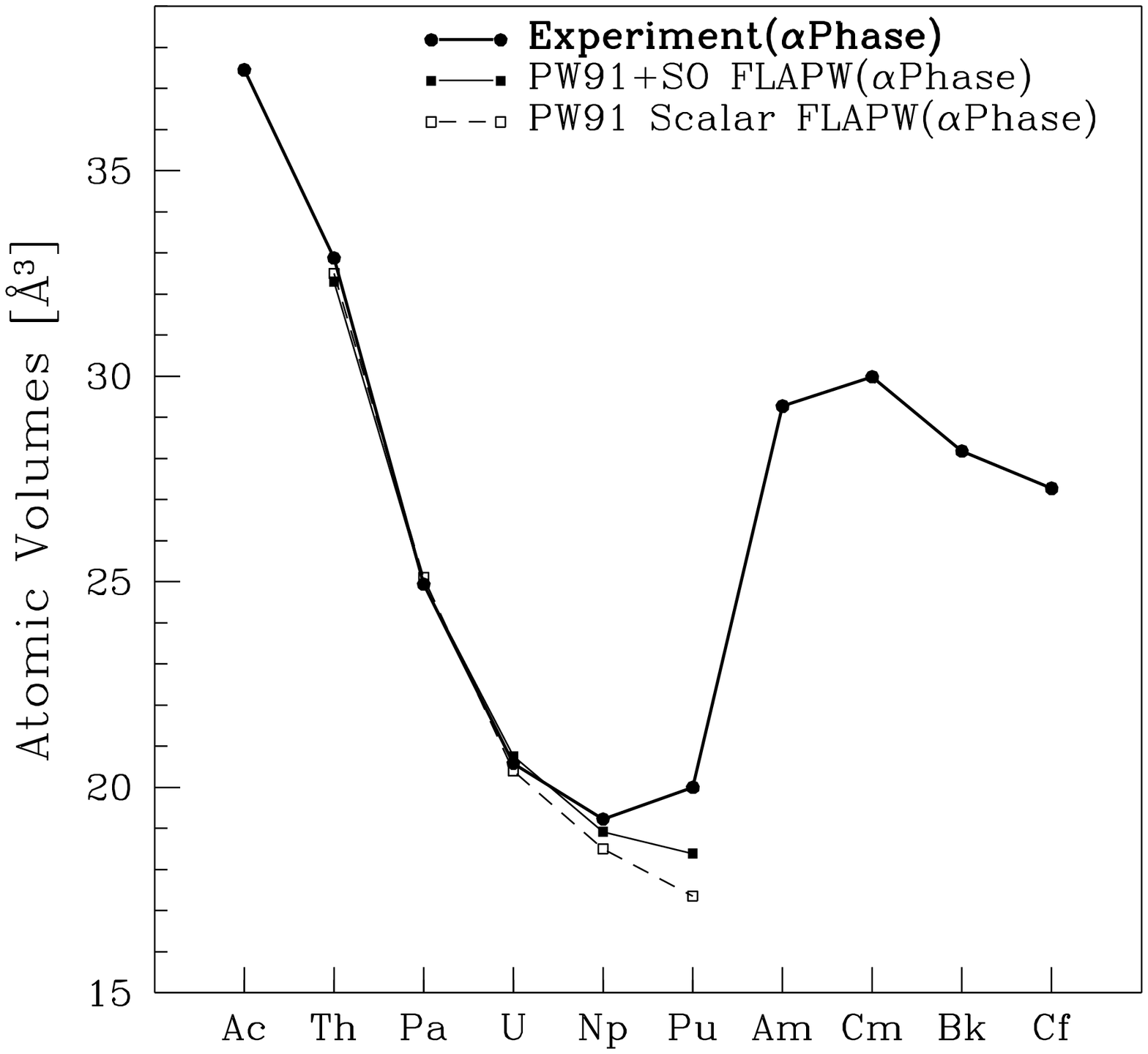,width=6.5in}
\caption{The equilibrium volume of the light actinides in the $\alpha$-phases
as calculated by FLAPW, both in the scalar-relativistic (open squares)
and fully-relativistic (solid squares) treatments.  These numerical results
were done using the GGA form of exchange and correlation.  Experimental
results are also shown (solid circles).}\label{fig:v0}
\end{figure}

\begin{figure}[htb]
\epsfig{file=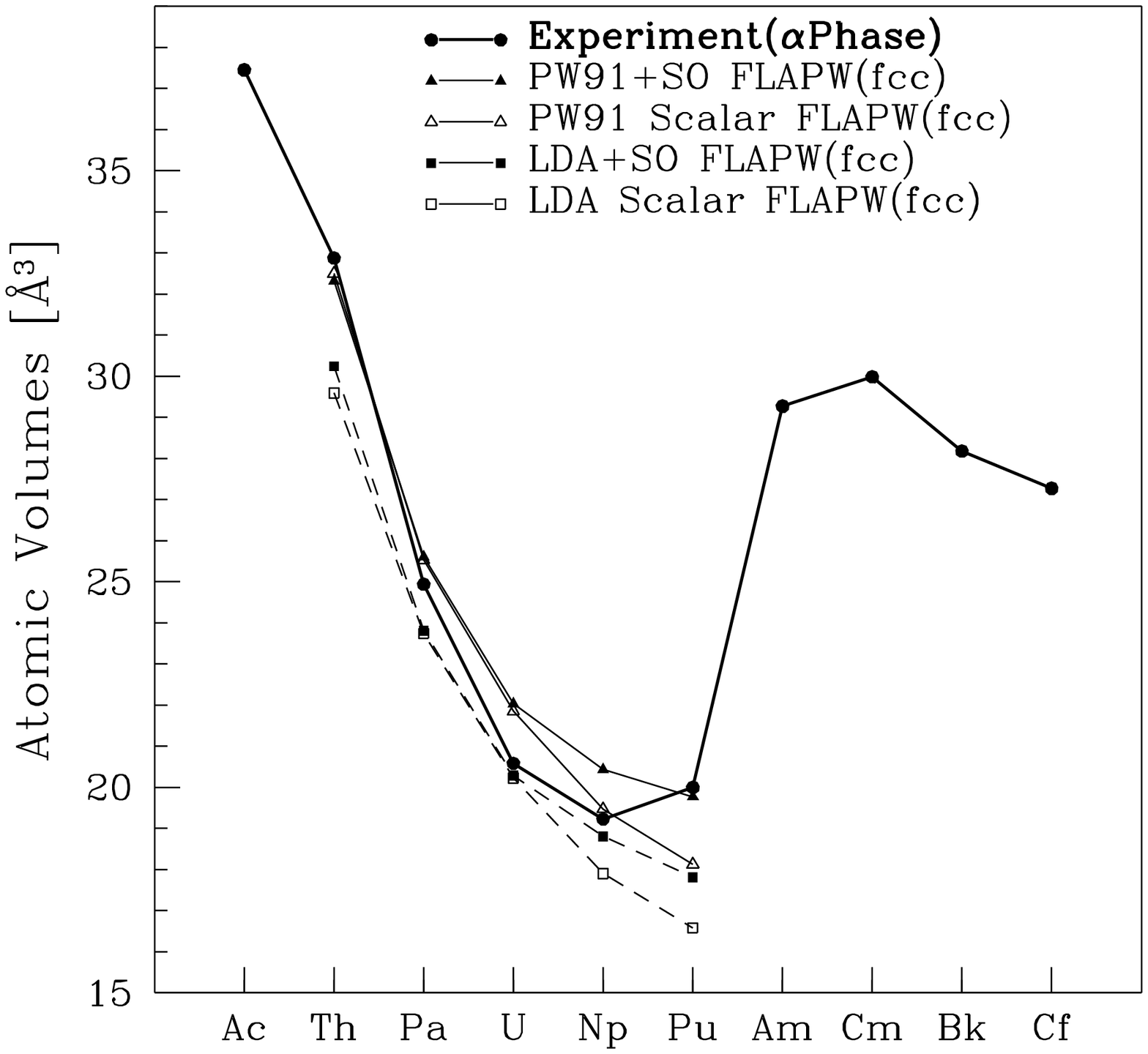,width=6.5in}
\caption{The equilibrium volume of the light actinides in the fcc phase
as calculated by FLAPW, both in the scalar-relativistic (open squares)
and fully-relativistic (solid squares) treatments.  These numerical results
were done using the GGA form of exchange and correlation.  Experimental
results are also shown (solid circles), but for the respective
$\alpha$ phases.}\label{fig:fcc_v0}
\end{figure}

The FLAPW and LCGTO-FF GGA results are compared with experiment (in
most cases only room temperature experimental data is available) in
Figure \ref{fig:v0} for the $\alpha$ phases.  The agreement between
theory and experiment is quite good.  A similar plot is shown in
Figure \ref{fig:fcc_v0} for the fcc crystal structures, in which we
see that the fcc have significant errors when compared to the $\alpha$
phase experimental volumes.  The fcc phases do, however, obey the same
general trend with regard to spin-orbit contributions; spin-orbit
effects have negligible impact on the equilibrium volumes for
Th$\rightarrow$U, and cause an expansion of the lattice for Np and Pu.
Similarly, GGA results are always expanded relative to LDA.

The upturn in equilibrium volume for Pu seen in Figure \ref{fig:v0}
has been a subject of some contention.  Based on LMTO-ASA studies,
this upturn was originally ascribed to spin-orbit effects on the
valence electrons.\cite{2,3} FP-LMTO studies by Wills and
Eriksson,\cite{4} however, not only did not find an upturn, but found
a contraction with the inclusion of spin-orbit coupling, leading them
to hypothesize that the LMTO-ASA calculations reached the wrong
conclusion about the effects of spin-orbit coupling due to their
treating the 6p semi-core states as core states.  Recently, however,
P\'enicaud\cite{penicaud} has performed a fully relativistic LMTO-ASA
calculation for the light actinides in which the 6p states were
treated as valence states (in a relativistic $j,\kappa$-basis), and
found no such contraction.  Indeed, the LMTO-ASA studies show a
systematic trend towards an expansion in volumes when spin-orbit
effects are treated perturbatively\cite{3} and
non-perturbatively.\cite{penicaud} Our present studies are consistent
with the LMTO-ASA results, and Figure \ref{fig:v0} shows that
spin-orbit effects on the valence electrons are responsible for a
gradual increase of the equilibrium volume as one proceeds along the
sequence of the light actinides, but it does not appear to account for
the anomalous upturn in volume for $\alpha$-Pu.

Another possibility for the upturn is that it represents a
finite-temperature effect (since experimental volumes are measured at
room temperature).  While it is true that thermal expansion increases
for Np and Pu, however, if the measured thermal expansion
coefficients\cite{gschneidner} (resulting in an increase for the
volume of 2.4\% for Np, and 4.8\% for Pu) are used to correct the
zero-temperature calculations, the two atomic volumes will be the same
at room temperature (about 130 in atomic units) and not have the
expected upturn.  It is quite interesting that our density-functional
calculations predict the atomic volumes so well until we reach Pu.
For fcc $\delta$-phase Pu, the theoretical volume is too small by
about 20\%.  Although this is an exceptionally large error for a GGA
calculation, it has been known for many years that $\delta$-Pu is
anomalous due to its position on the boundary between the light
actinides that have itinerant 5$f$-electrons and the heavy actinides
that have localized 5$f$-electrons (which form a second rare-earth
like series).  Although the volume found here for $\delta$-Pu is 10\%
larger than that found with FP-LMTO method,\cite{4} that increase is
not large enough to remove the discrepancy between theory and
experiment.  However, the large experimental difference in volume
between the low-temperature $\alpha$ and much higher temperature
$\delta$ phases suggest that new physics (beyond LDA or GGA) is
responsible for the large expansion of the $\delta$ phase (the
enormous difference cannot be explained by any known thermal expansion
mechanism).  It is likely that large intra-atomic electron-electron
coulomb correlations (a large effective Hubbard U) are responsible for
the anomalous properties of $\delta$-Pu.  It may even be possible that
some localization is already apparent in the $\alpha$ phase for Pu,
thus accounting for the slight (around 5\%) discrepancy between the
computed and measured atomic volume, and causing the upturn.  In heavy
fermion systems, it has been argued\cite{zwicknagl} that the effects
of strong electron-electron correlations (which is responsible for the
huge specific-heat enhancements) leads to heavier band masses
(renormalized bands with a reduced effective band width).  In terms of
a Friedel model description of bonding,\cite{friedel} such an effect
(a narrower $f$ band width) would reduce the strength of the $f$
bonding, which should expand the lattice.

\begin{figure}[htb]
\epsfig{file=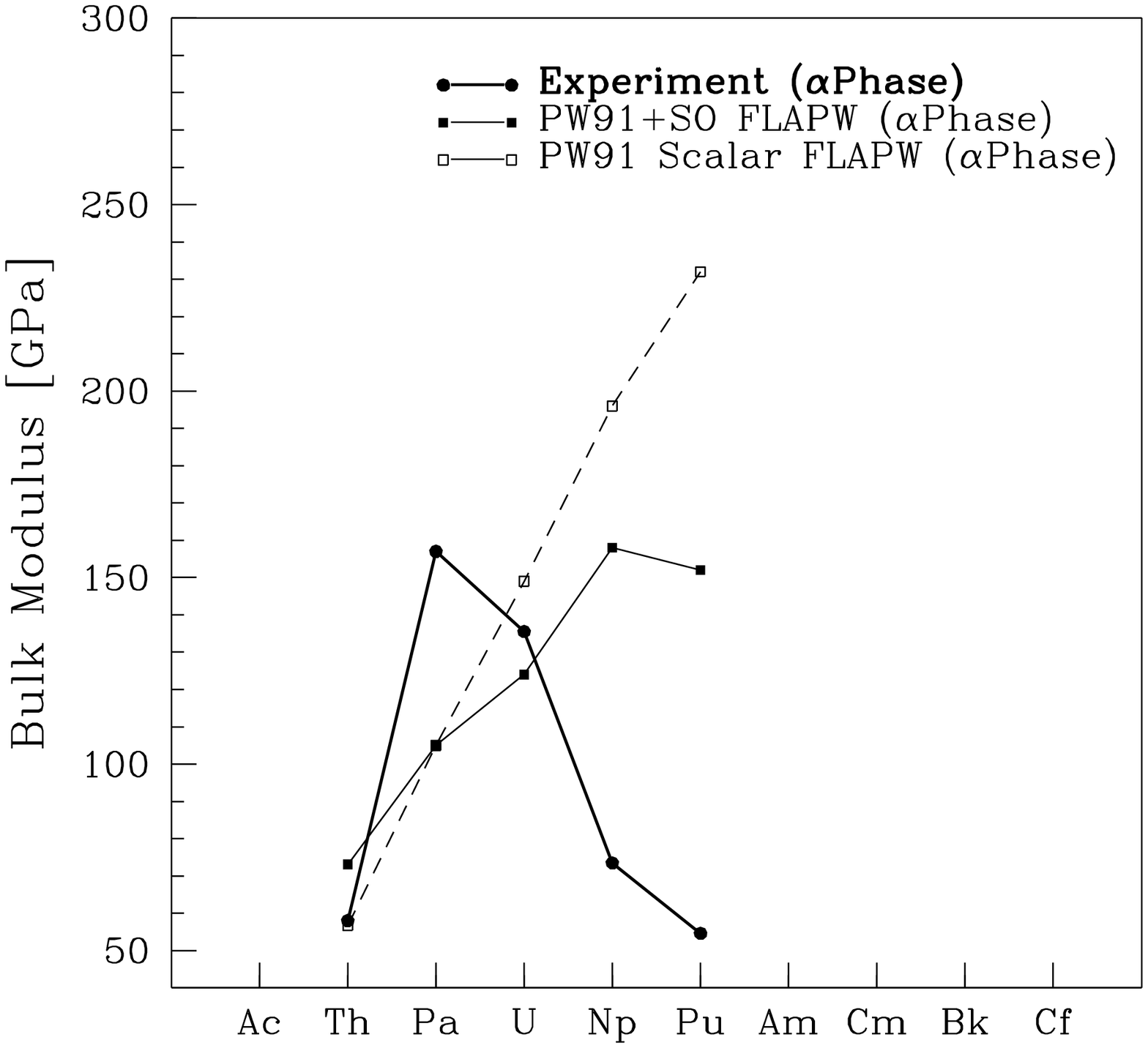,width=6.5in}
\caption{The bulk moduli of the light actinides in the $\alpha$ phases,
as calculated by FLAPW, both in the scalar-relativistic (open squares)
and fully-relativistic (solid squares) treatments within the GGA, compared 
to experiment (solid circles).}\label{fig:B0}
\end{figure}

Table \ref{tbl_B0} lists our results for the bulk modulus for the
light actinides Th$\rightarrow$Pu, in both the fcc and $\alpha$
phases.  Since the FP-LMTO calculations found a substantial decrease
in the equilibrium volumes, it not very surprising that they predict a
correspondingly larger bulk modulus.  Comparison with the
experimentally determined bulk moduli, however, which is shown in
Figure \ref{fig:B0}, show that our most accurate fully relativistic
GGA FLAPW results are in somewhat strong disagreement, particularly
for the heavier actinides, Np and Pu.  We do not recover the turnover
seen experimentally past Pa.  There are several possible reasons for
this discrepancy.  One is that the bulk modulus is more sensitive than
the equilibrium volume to numerical errors in our calculations.
Another explanation is that temperature effects (all of the available
experimental data is at room temperature or above) may play an
increasingly important role, softening the bulk modulus, especially
for Np and Pu.  Strong anharmonic effects have been found recently in
the Debye-Waller temperatures for U, Np, and Pu.\cite{lawson}

\section{Conclusions}
\label{sec:conc}
In summary, it has been demonstrated that the LCGTO-FF and FLAPW
electronic-structure methods produce a zero-pressure volume and bulk
modulus for the light actinides Th$\rightarrow$Pu that are in good
agreement with each other but differ significantly from previous
full-potential results obtained from the FP-LMTO method,\cite{4} with
the present volumes being roughly 3-10\% larger.  Given the good
agreement between the LCGTO-FF and FLAPW methods, it seems likely that
the earlier FP-LMTO results were in error.  If so, this might account
for the 3\% to 10\% underestimate of the volumes found for all of the
light-actinide crystals with the FP-LMTO method using the GGA
exchange-correlation potentials.  The present results do not affect
the long standing problem of the anomalous volume of $\delta$-Pu,
which presumably is due to correlation effects that are not adequately
accounted for in either the LDA or the GGA.

\acknowledgments

JCB thanks J. P. Perdew for providing the subroutines used to
implement the PW91 GGA in GTOFF.  This research is supported by the
Department of Energy under contract W-7405-ENG-36.  This research used
resources of the National Energy Research Scientific Computing Center,
which is supported by the Office of Energy Research of the
U.S. Department of Energy under Contract No. DE-AC03-76SF00098.

% tables 
%
\newpage
\begin{table}
% \sqeezetable
\caption{The equilibrium volume (atomic units) obtained here for the light 
actinides with 
LCGTO-FF and FLAPW calculations using the LDA and GGA models, with and
without spin-orbit (SO) effects included, are compared to results from
LMTO-ASA calculations with combined correction terms and calculated 
temperature expansion corrections,\cite{3} FCD-LMTO
calculations,\cite{5} and FP-LMTO calculations.\cite{4}}
\label{tbl_V}
\begin{tabular}{lllllllll}
           &\multicolumn{4}{c}{cF4 (fcc)}&\multicolumn{4}{c}{$\alpha$-phase}\\
Method     & LDA & +SO & GGA & +SO & LDA & +SO & GGA & +SO \\
\cline{2-5}\cline{6-9}\\
%\cline{1-6}
%\multicolumn{6}{c}{Ac}\\
%FLAPW      &&  272.57   &          &  312.88  &          \\
%Experiment & 252.73 \\
%\cline{1-9}
&\multicolumn{8}{c}{\bf Th}\\
%\cline{1-9}
&\multicolumn{4}{c}{cF4}\\
\cline{2-5}\\
LMTO-ASA   & 232.1 & 229.4 &        &         \\
FCD-LMTO   & 212.7 &       & 233.7 &         \\
FP-LMTO    &       & 182.1 &        & 199.9  \\
LCGTO-FF   & 200.5 & 200.4 & 216.9 & 216.2  \\
FLAPW      & 199.7 & 204.1 & 219.3 & 218.1  \\
Experiment &\multicolumn{4}{c}{221.7(298K)\tablenote{Reference [20].}} \\
%\cline{1-9}
&\multicolumn{8}{c}{\bf Pa}\\
%\cline{1-9}
&\multicolumn{4}{c}{cF4}&\multicolumn{4}{c}{tI2}\\
\cline{2-5}\cline{6-9}\\
LMTO-ASA   &       &       &       &       & 178.8 & 177.5 &       &       \\
FCD-LMTO   & 171.7 &       & 182.8 &       & 164.8 &       & 175.8 &       \\ 
FP-LMTO    &       & 149.9 &       & 160.2 &       & 147.7 &       & 157.6 \\
LCGTO-FF   & 159.6 & 160.6 & 171.5 & 172.8 & 155.9 &       & 168.7 & 170.4 \\
FLAPW      & 160.2 & 160.7 & 172.3 & 172.8 & 156.8 & 155.8 & 169.3 & 168.4 \\
Experiment &       &       &       &       &\multicolumn{4}{c}{168.30(298K)\tablenote{Reference [21].}} \\
%\cline{1-9}
&\multicolumn{8}{c}{\bf U}\\
%\cline{1-9}
&\multicolumn{4}{c}{cF4}&\multicolumn{4}{c}{oC4}\\
\cline{2-5}\cline{6-9}\\
LMTO-ASA   &       &       &       &       & 144.4 & 147.1 &       &      \\
FCD-LMTO   & 147.0 &       & 157.1 &       & 131.4 &       & 141.2 &  \\
FP-LMTO    &       & 129.4 &       & 138.6 &       & 123.7 &       & 131.5  \\
LCGTO-FF   & 136.1 & 138.8 & 146.1 & 149.7 & 128.9 &       & 138.4 & 139.7  \\
FLAPW      & 136.5 & 136.9 & 147.5 & 148.7 & 127.9 & 128.5 & 137.7 & 140.1  \\
Experiment &       &       &       &       &\multicolumn{4}{c}{138.89(4.2K)\tablenote{Reference [22].}}\\
%\cline{1-9}
&\multicolumn{8}{c}{\bf Np}\\
%\cline{1-9}
&\multicolumn{4}{c}{cF4}&\multicolumn{4}{c}{oP8}\\
\cline{2-5}\cline{6-9}\\
LMTO-ASA   &       &       &       &       & 125.5 & 128.2 &       &       \\
FCD-LMTO   & 137.4 &       & 144.1 &       & 117.8 &       & 131.8 &  \\
FP-LMTO    &       & 116.7 &       & 125.8 &       & 112.0 &       & 122.1,
124.2\tablenote{Reference [33].}  \\
LCGTO-FF   & 121.1 & 125.7 & 131.6 & 138.2 & \\
FLAPW      & 120.8 & 126.9 & 131.4 & 137.9 & 116.4 & 118.3 & 124.6 & 127.7  \\
Experiment &       &       &       &       &\multicolumn{4}{c}{129.9(293K)\tablenote{Reference [23].}}\\
%\cline{1-9}
&\multicolumn{8}{c}{\bf Pu}\\
%\cline{1-9}
&\multicolumn{4}{c}{cF4}&\multicolumn{4}{c}{mP16}\\
\cline{2-5}\cline{6-9}\\
LMTO-ASA   & 114.7 & 130.9 &       &       & 114.7 & 130.9 &       &      \\
FCD-LMTO   & 132.8 &       & 140.0 &       & 112.2 &       & 131.4 &  \\
FP-LMTO    &       & 109.2 &       & 119.2 &       &       &      & 124.2$^{\rm d}$\\
LCGTO-FF   & 111.5 & 119.8 & 121.2 & 134.4  \\
FLAPW      & 111.9 & 120.2 & 122.3 & 133.4 & 109.7 & 114.2 & 117.2 & 124.4 \\
Experiment &\multicolumn{4}{c}{168.0(653K)\tablenote{Reference [24].}}
&\multicolumn{4}{c}{134.95(294 K)\tablenote{Reference [25].}} \\
\end{tabular}
\end{table}

\begin{table}
% \sqeezetable
\caption{The bulk modulus (GPa) obtained here for Th$\rightarrow$Pu with 
LCGTO-FF and 
FLAPW calculations using the LDA and GGA models, with and without
spin-orbit (SO) effects included, are compared to results from FP-LMTO
calculations\cite{4} and room temperature data.}
\label{tbl_B0}
\begin{tabular}{lrrrrrrrr}
           &\multicolumn{4}{c}{cF4 (fcc)}&\multicolumn{4}{c}{$\alpha$-phase}\\
Method     & LDA & +SO & GGA & +SO & LDA & +SO & GGA & +SO \\
\cline{2-5}\cline{6-9}\\
%
%\multicolumn{6}{c}{Ac}\\
%FLAPW      &   34.1    &          &   24.0   &          \\
%Experiment & 25.0 \\
%\cline{1-9}
&\multicolumn{8}{c}{\bf Th}\\
%\cline{1-9}
&\multicolumn{4}{c}{cF4}\\
\cline{2-5}\\
FP-LMTO    &      & 82.6 &      & 61.5   \\
LCGTO-FF   & 64.8 & 68.7 & 58.8 & 61.7   \\
FLAPW      & 61.0 & 78.6 & 56.7 & 73.1   \\
Experiment &\multicolumn{4}{c}{58(1)\tablenote{Reference [26].}} \\
%\cline{1-9}
&\multicolumn{8}{c}{\bf Pa}\\
%\cline{1-9}
&\multicolumn{4}{c}{cF4}&\multicolumn{4}{c}{tI2}\\
\cline{2-5}\cline{6-9}\\
FP-LMTO    &     & 141 &     & 122 &     & 146 &     & 123 \\
LCGTO-FF   & 121 & 120 & 102 & 102 & 112 &     &  96 &  98 \\
FLAPW      & 123 & 104 & 100 &  96 & 111 & 110 & 105 & 105 \\
Experiment &     &     &     &     &\multicolumn{4}{c}{157(5)\tablenote{Reference [27].} }\\
%\cline{1-9}
&\multicolumn{8}{c}{\bf U}\\
%\cline{1-9}
&\multicolumn{4}{c}{cF4}&\multicolumn{4}{c}{oC4}\\
\cline{2-5}\cline{6-9}\\
FP-LMTO    &     & 186 &     & 148 &     & 240 &     & 172  \\
LCGTO-FF   & 160 & 150 & 101 & 104 & 181 &     & 138 & 126  \\
FLAPW      & 148 & 228 & 125 &  99 & 176 & 144 & 149 & 124  \\ 
Experiment &     &     &     &     &\multicolumn{4}{c}{135.5\tablenote{Reference [28].}} \\
%\cline{1-9}
&\multicolumn{8}{c}{\bf Np}\\
%\cline{1-9}
&\multicolumn{4}{c}{cF4}&\multicolumn{4}{c}{oP8}\\
\cline{2-5}\cline{6-9}\\
FP-LMTO    &     & 199 &     & 161 &     & 300 &     & 170  \\
LCGTO-FF   & 185 & 158 & 142 & 121 & \\
FLAPW      & 190 & 136 & 137 & 140 & 260 & 234 & 196 & 158  \\
Experiment &     &     &     &     &\multicolumn{4}{c}{73.5\tablenote{Reference [29].}}\\
%\cline{1-9}
&\multicolumn{8}{c}{\bf Pu}\\
%\cline{1-9}
&\multicolumn{4}{c}{cF4}&\multicolumn{4}{c}{mP16}\\
\cline{2-5}\cline{6-9}\\
FP-LMTO    &     & 214 &     & 143 &     &     &     & 130\tablenote{Reference [33].}   \\
LCGTO-FF   & 218 & 130 & 170 & 102    \\
FLAPW      & 194 & 143 & 153 & 121 & 307 & 244 & 232 & 153  \\
Experiment &     &     &     &     &\multicolumn{4}{c}{47.2(7)\tablenote{Reference [30].}, 
54.6\tablenote{Reference [31].}}\\
\end{tabular}
\end{table}

\begin{figure}[htb]
\epsfig{file=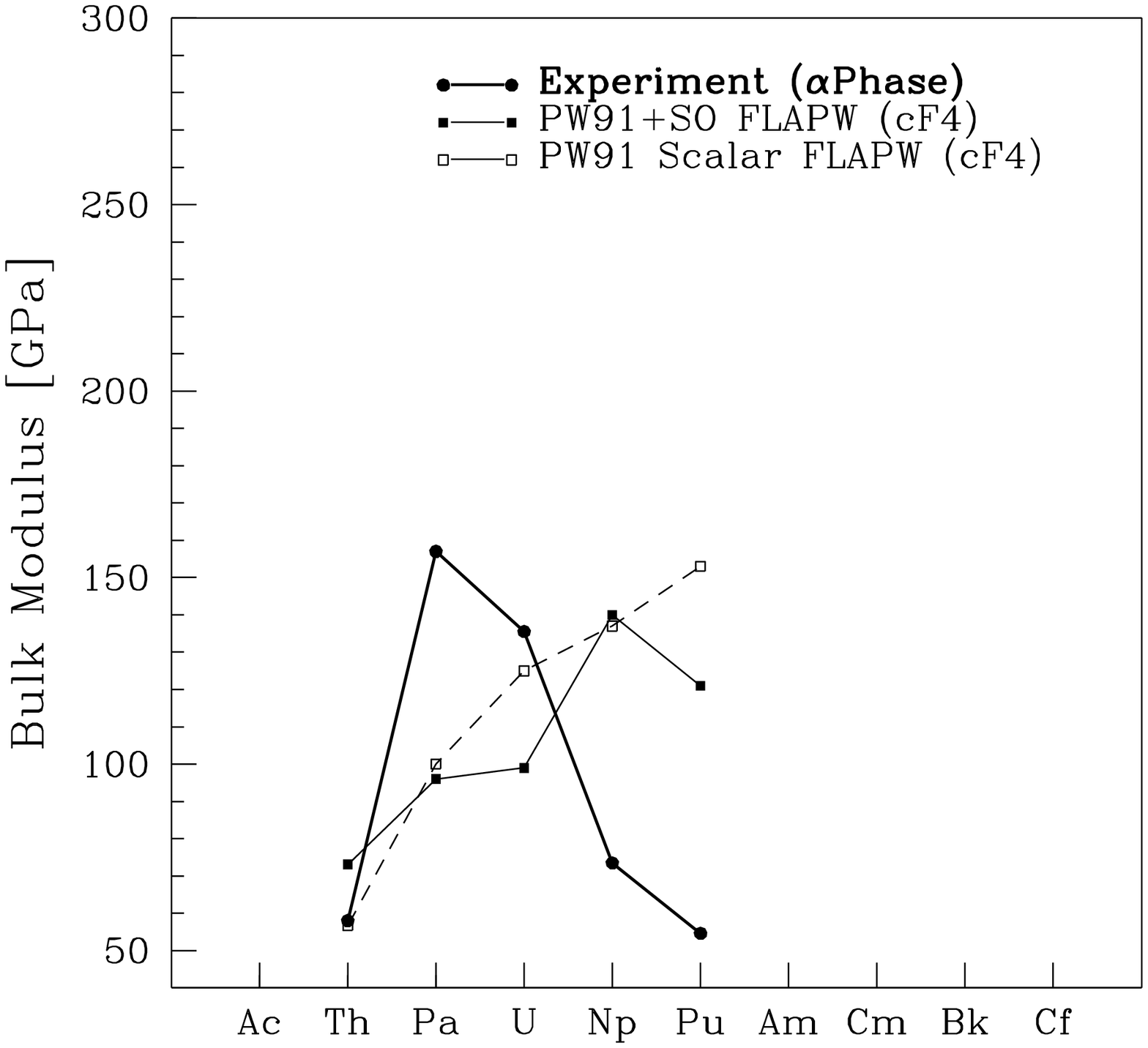,width=6.5in}
\caption{The bulk moduli of the light actinides in the fcc phase,
as calculated by FLAPW, both in the scalar-relativistic (open squares)
and fully-relativistic (solid squares) treatments within the GGA.
For comparison, these values are compared to experiment (solid circles)
in the respective $\alpha$ phases.}\label{fig:fccB0}
\end{figure}

\end{document}